\documentclass{appolb}

\usepackage{epsfig}
\usepackage{amsmath}
\usepackage{amssymb}

\newcounter{figures}
\newcounter{tables}
\begin{document}
\title{Is there a non-standard-model contribution in\\ non-leptonic $b\to s$ decays?%
\thanks{Presented at the FlaviaNet workshop ``Low energy constraints on extensions of the Standard Model,
23-27 July 2009, Kazimierz, Poland}%
}
\author{Martin Jung
\address{Instituto de F\'{\i}sica Corpuscular (IFIC),
CSIC-Universitat de Val\`encia, \\
Apartado de Correos 22085,
E-46071 Valencia, Spain}
}
\maketitle
\begin{abstract}
The data on high-pre\-ci\-sion flavour ob\-ser\-vables reveal certain
\emph{puzzles} when compared to Standard Model expectations based on a global fit of the CKM unitarity triangle and general
theoretical estimates. The discussion of these tensions in the channels $B\to J/\psi K$, $B\to\phi K$, and $B\to\pi K$, 
and the deduced constraints for New Physics operators of the class $b\to s\bar{q}q$  form the content of this talk.
\end{abstract}

\section{Introduction}
$b\to s$ transitions tend to be a good ground for NP searches,
due to the hierarchy in the relevant CKM matrix elements.
Of these, three groups of non-leptonic decays are discussed: $B\to J/\psi K$,  $B\to \phi K$, and  $B\to \pi K$. All of them are 
``puzzling'', i.e. tensions with the SM expectations are found, and the data for these decays are relatively precise.
This motivates the introduction of NP contributions by operators of the form $\mathcal{O}^{b\to s}_q=(\bar{s}b)(\bar{q}q)$.
The analysis presented here follows \cite{FJM}, updating and slightly enhancing the analysis performed there.

In the following the assumptions will be made, that one  operator dominates the NP contributions, leading to a single weak phase 
for the corresponding matrix elements, while meson mixing is unaffected. The colour and Dirac structure of the operators will not be specified.

\section{Unitarity triangle analysis}
As a first step, the CKM parameters entering the analysis have to be determined in an independent way. This is done for the considered scenario
using the input from semileptonic decays and $B_{d,s}$-mixing, only. The values used in the analysis are the averages performed by the 
CKMfitter group \cite{CKMfitter}, as presented on the conference in Moriond 2009, leading to $\sin2\beta=0.746 ^{+0.014}_{-0.020} \pm 0.081$
and $\gamma=(65.7^{+1.8}_{-1.7}\pm 5.5)^\circ$, where the first error is treated as gaussian, the second as flat.
Note that, while the constraint from $BR(B\to\tau\nu)$ is not used in the following, its inclusion would strongly enhance the 
slight tension visible here, dependent however on the determination of $f_B$.
%
%
%
%
%
\section{$\mathbf{B\to J/\boldsymbol{\psi} K}$\label{NPinbtosAmp}}
This decay, often referred to as the \emph{Golden Mode}, plays a special role in the SM, because it is dominated to very good 
approximation by only one isospin amplitude, 
leading to $S_{J/\psi K}=\sin2\beta$, and critical observables \cite{FleischerMannel1} $\Delta A_{\rm CP}=0$ and $A_I=0$.
Regarding corrections from subleading operators, see \cite{Boos,Li,GronauRosnersuppressedterms,Ciuchini2005,FFJM}.
From the data, small deviations from this pattern are observed, see the winter 2009 averages in \cite{HFAG}.
Turning now to the hypothesis of NP in the decay amplitudes as described above, the general parametrization reads
\begin{equation}\label{ParametrizationJPsiK}
{\cal A} (\bar{B}^{0,-} \to J/\psi \bar K^{0,-}) 
 = {\cal A}_0 \left[ 1 + r_0 \, e^{i \theta_W}  e^{i \phi_{0}} \pm
    r_1 \, e^{i \theta_W}  e^{i \phi_{1}} \right] \,,
\end{equation}
with $r_{0,1},\,\phi_{0,1}$ and $\theta_W$ denoting the moduli, strong and weak phases of the NP amplitudes with $\Delta I=0,1$ respectively.

In order to keep track of the different effects determining the order of magnitude for different contributions, a
power-counting is introduced \cite{FleischerMannel1,Gronauetal1}, combining the Wolfenstein hierarchy, (electroweak) penguin suppression factors ($\mathcal{O}(\lambda(\lambda^2))$) and an estimate of the ``generic size'' of NP contributions
$\mathcal{A}_{\rm NP}\sim M_W^2/\Lambda_{\rm NP}^2\langle\mathcal{O}_{\rm NP}\rangle\sim \lambda\times \mathcal{A}_0\,.$

\begin{figure}[t!!bt]
\begin{center}
\parbox{0.45\textwidth}{
\rotatebox{90}{\hspace{0.16\textwidth}\footnotesize $r_0 \, \sin\phi_0$}
\includegraphics[width=0.34\textwidth]{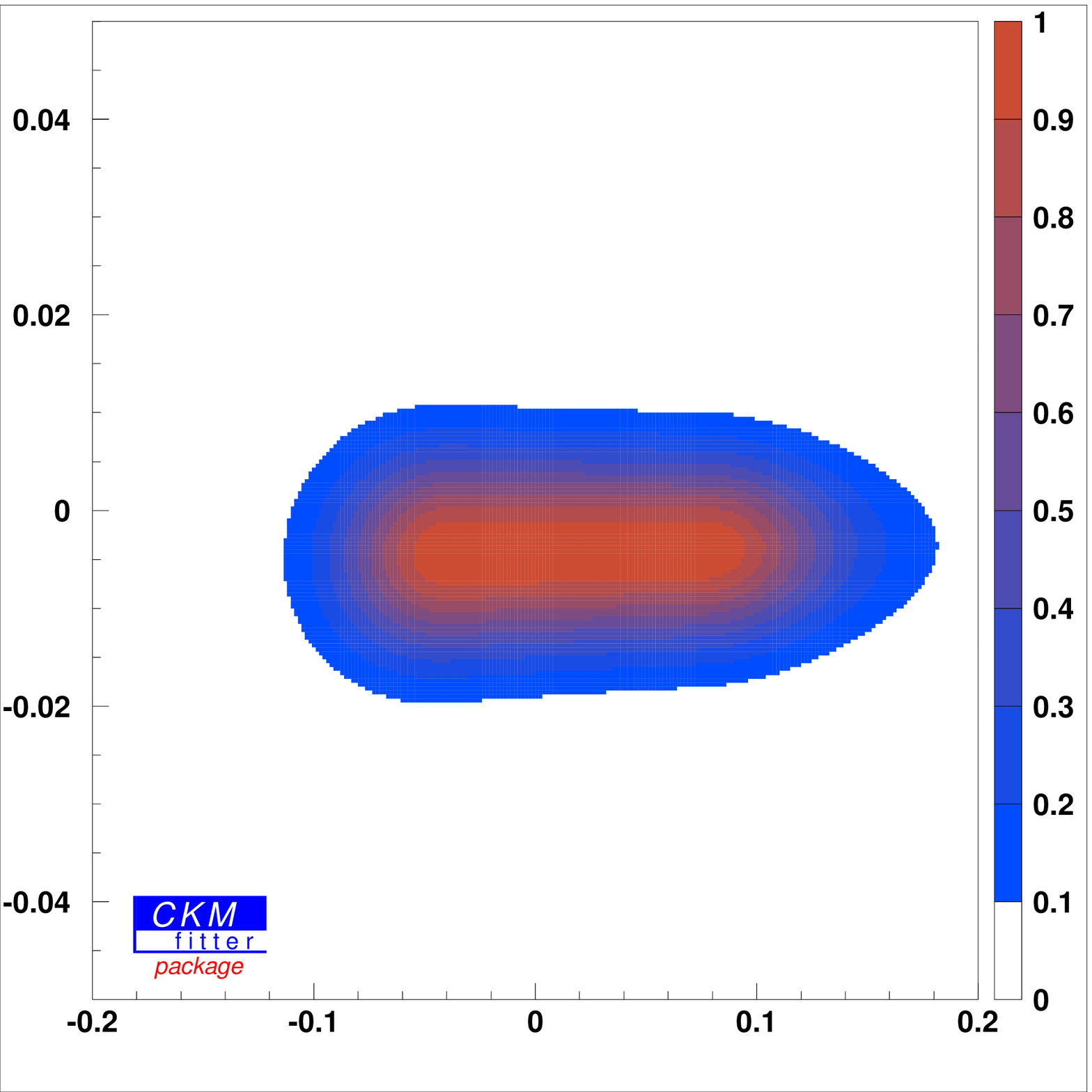}\\
\centerline{\footnotesize   $r_0 \, \cos\phi_0$}}
\quad
\parbox{0.45\textwidth}{
\rotatebox{90}{\hspace{0.16\textwidth}\footnotesize $r_1 \, \sin\phi_1$}
\includegraphics[width=0.34\textwidth]{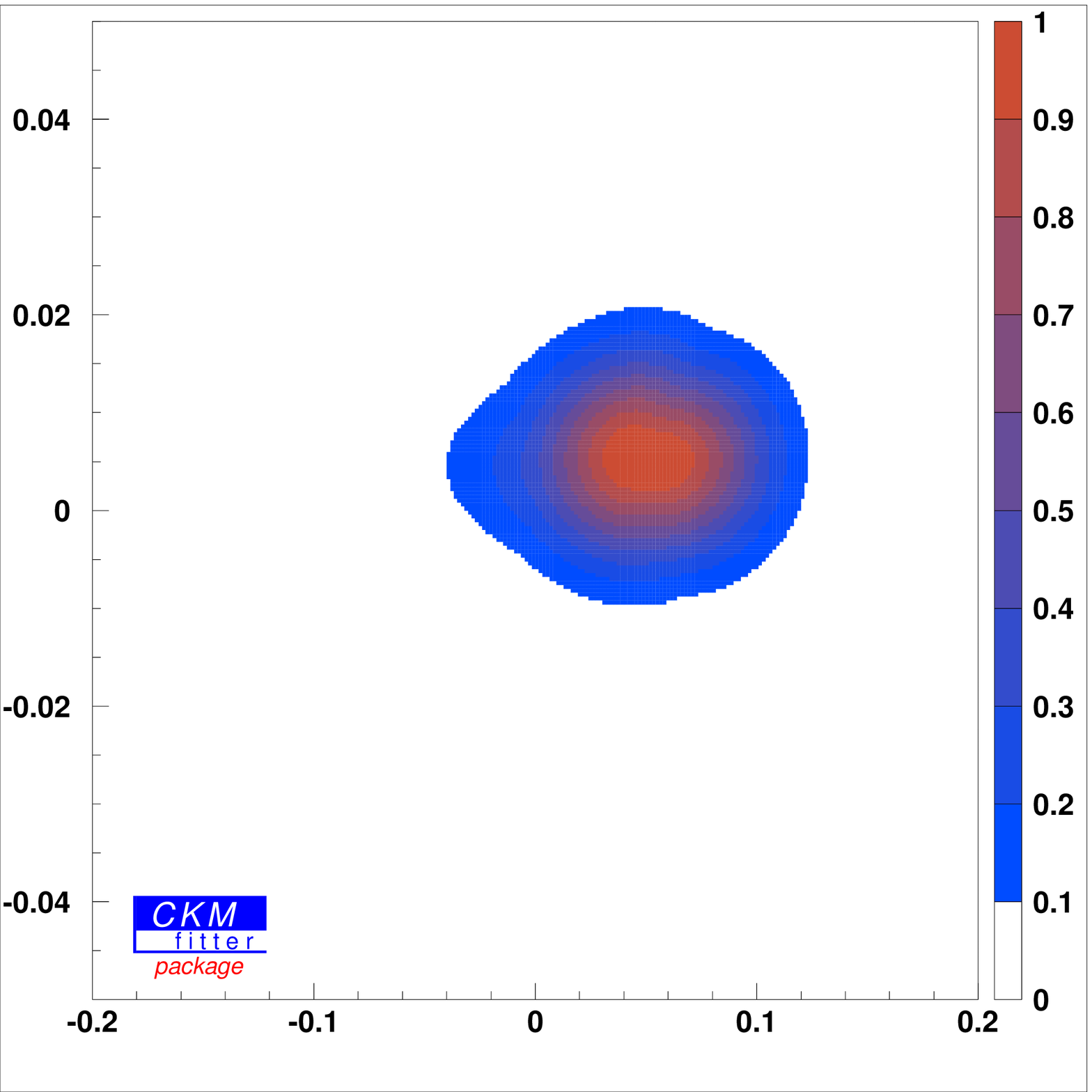}\\
\centerline{\footnotesize  $r_1 \, \cos\phi_1$
}}
\end{center}\vspace{-2ex}
\caption{\label{fig:resultsJPsiKI01}\footnotesize{The result for  $r_0 \, e^{i\phi_0}$ and $r_1 \, e^{i \phi_1}$ 
from the fit to $J/\psi K$ observables, see text.
}}
\end{figure}

Taking the data at face value, the observed $A_I\neq 0,\,\Delta A_{\rm CP}\neq 0$ imply a $\Delta I=1$ amplitude with a new weak phase, stemming from an operator $(\bar{s}b)(\bar{u}u/\bar{d}d)$.
For fits with $\Delta I=0$, only, see \cite{FJM,JPhd}. In the following, the weak phase is 
set to $\pi-\gamma$ for simplicity, the solutions for other values of $\theta_W$ can be obtained from reparametrization invariance, see \cite{FJM}.
One can trivially fit all observables. The fit result is 
plotted in figure~\ref{fig:resultsJPsiKI01}. The $1\sigma$ parameter ranges are given by
$r_0 \, \cos\phi_0 = \left[-0.074,\, 0.118 \right]$, $r_0 \, \sin\phi_0 = \left[-0.015,\,0.003 \right]$,
$r_1 \, \cos\phi_1 = \left[\phantom{-}0.014,\,0.089 \right]$ and $r_1 \, \sin\phi_1 = \left[-0.002,\,0.013 \right]$.
The fitted parameters have reasonable orders of magnitude, although generally $r_1\ll r_0$ is expected.
Notice that the preferred values for the strong phases turn out to be small. 
Notice furthermore that, depending on the actual size of these suppression 
factors, the result for $r_0$ and $r_1$ may also be interpreted as due to unexpectedly large effects from subleading SM operators.
%
%
\section{$\mathbf{B\to \boldsymbol{\phi} K}$}
The similar analysis for this penguin-dominated decay results in 
a power-counting $A_{\rm NP}(\Delta I=0)/(\mathcal{A}_0)\lesssim\mathcal{O}(1)\,,A_{\rm NP}(\Delta I=1)/(\mathcal{A}_0)\lesssim\mathcal{O}(\lambda)\,.$
SM estimates usually give small subleading contributions \cite{Grossmanetal,WilliamsonZupan,Chengetal,Beneke}. 
The parametrization is completely analogous to the one of $B\to J/\psi K$. Again, tensions with the naive SM expectations are found \cite{HFAG}.
Again only the results including isospin breaking contributions are shown.
Note that since the publication of \cite{FJM}, the data of the time-dependent CP asymmetries changed significantly. 
The corresponding fit yields the $1\sigma$-ranges $r_0 \, \cos\phi_0 = \left[\phantom{-}0.03,\,0.48 \right]$, $r_0 \, \sin\phi_0 = \left[-0.11,\,-0.03 \right]$ , $r_1 \, \cos\phi_1 = \left[-0.35,\,0.10 \right]$ and $r_1 \, \sin\phi_1 = \left[-0.09,\,-0.01 \right] \,,$
favouring larger values than in $B\to J/\psi K$. Also in this case small phases are preferred. In addition, the fit yields non-vanishing values for both contributions, with the contribution to $\Delta I=0$ tending to be larger. 
Importantly, also here an operator with the structure $(\bar{s}b)(\bar{u}u/\bar{d}d)$ is needed to explain all deviations, and the 
relative size of the effects in $B\to J/\psi K$ and $B\to \phi K$ corresponds to naive expectations, when assigning the deviations 
to the same source.
%
\section[$B\to \pi K$]{$\mathbf{B\to \boldsymbol{\pi} K}$}
$B\to\pi K$ decays are also penguin dominated, due to the Cabbibo suppression of their tree contributions. In addition, they are sensitive
to electroweak penguin contributions.
In the following, the parametrization from \cite{Neubert1998,BBNS1} is used for the decay amplitudes.
The experimental data is given in table~\ref{tab:bestfit}. Without any assumptions on strong interaction dynamics, in the isospin limit 
one is left with 11 independent hadronic parameters for 9 observables. 
In order to test the SM against possible NP effects in these decays, one needs therefore additional dynamical input, implying a stronger
model dependence.
The following assumptions are used here:
The tiny doubly Cabbibo suppressed penguin contribution ($\epsilon_a$) is set to zero, and the values from \cite{BBNS1} 
for $q,q_C$ and the corresponding phases are used.
Tensions in the fit, or incompatible values for the parameters $\epsilon_{T,3/2}$ and $\phi_{T,3/2}$ then may be taken 
as indication for possible NP contributions.
\begin{table}[t!bt!]
  \begin{center}\small{
  \begin{tabular}{|c|c||c|c|}
  \hline Observable & HFAG \cite{HFAG} & SM fit
  & NP $(I=0,1)$
  \\
  \hline
  $\overline{{\rm BR}}(\pi^0K^-) \cdot 10^{6}$            & $12.9 \pm 0.6$
  & $12.4$ & $12.8$
  \\
  $\overline{{\rm BR}}(\pi^-\bar{K}^0)\cdot 10^{6}$       & $23.1 \pm 1.0$  &
  $23.7$ & $23.3$
  \\
  $\overline{{\rm BR}}(\pi^+K^-)\cdot 10^{6}$       & $19.4 \pm 0.6$  &
  $19.7$  & $19.5$
  \\
  $\overline{{\rm BR}}(\pi^0\bar{K}^0)\cdot 10^{6}$ & $\phantom{1}9.8 \pm 0.6$  &
  $\phantom{1}9.3$  & $\phantom{1}9.7$
  \\\hline
  $\mathcal{A}_{\rm CP}(\pi^-\bar{K}^0)$          & $\phantom{-}0.009\pm0.025$ &
  $0^*$ & $0^*$
  \\
  $\mathcal{A}_{\rm CP}(\pi^0K^-)$                & $\phantom{-}0.050\pm0.025$ &
  $\phantom{-}0.043$ & $\phantom{-}0.047$
  \\
  $\mathcal{A}_{\rm CP}(\pi^+K^-)$                & $-0.098^{+0.012}_{-0.011}$       &
  $-0.098$ & $-0.092$
  \\\hline
  $ \eta_{\rm CP} \, S_{\pi^0 K_S}$                            & $-0.57\pm0.17$         &
  $-0.62$  & $-0.78$
  \\
  $C_{\pi^0K_S}$                            & $\phantom{-}0.01\pm0.10$
  & $\phantom{-}0.14$ & $\phantom{-}0.10$
  \\\hline 
\end{tabular}}
\end{center}\vspace{-2.5ex}
\caption{\footnotesize{\label{tab:bestfit}Experimental data for $B\to \pi K$ decays vs.\ various best fit results, see also text.
$\chi^2_{\rm SM}/{\rm d.o.f.}=3.8/3$ and 
$\chi^2_{\rm \pi-\gamma}/{\rm d.o.f.}=2.6/3$.
  }}
  \end{table}
The best fit values are shown in  in table~\ref{tab:bestfit}, showing clearly the reduction of $B\to\pi K$ puzzle for the new data. 
Especially the key parameter $\Delta\epsilon$ now corresponds to $|\Delta\epsilon/\epsilon_T|=|C/T|\in[0.22,1.00] (1\sigma)\,,$
which does not seem unreasonable. This led the authors of \cite{Ciuchinietal2008} to the conclusion that 
the data are now compatible with the SM. On the other hand, in another paper \cite{Fleischeretal2008} it has been concluded 
that the pattern of the measured time-dependent CP asymmetries shows a tension with the values predicted from $B^-\to\pi^-\pi^0$ 
with aid of $SU(3)$ arguments (fixing mainly $\epsilon_{3/2}$), leading to $S_{\pi^0K_S}\sim1$ and $C_{\pi^0K_S}=\Delta A$, and 
hinting towards a  electroweak penguin sector with a large weak phase.
In \cite{GronauRosner2008} an analysis along similar lines was performed, pointing out that (i) $C_{\pi^0K_S}\simeq\Delta A$ is an 
approximate result of a model-independent sumrule \cite{Gronau2005}, holding at the percent level, 
and (ii) that $S_{\pi^0K_S}\sim1$ is extremely sensitive to ${\rm BR}(B\to\pi^0K^0)$.
Finally, the authors of \cite{Baeketal2009} find a reduced puzzle, using the Neubert-Rosner relation for $q$ and its counterpart for colour-suppressed penguins. They find the tension not significantly relaxed by introducing modified electroweak penguins. 
This model-dependence clearly has to be clarified before any reliable conclusions are possible. The value for $\Delta A$ still implies large non-factorizable contributions, when interpreted in SM terms. In addition, the possible effects in $B\to J/\psi K$ and $B\to\phi K$ data 
should have an even more pronounced effect in $B\to\pi K$. This motivates the 
inclusion of NP operators along similar lines as in $B\to J/\psi K$ and $B\to\phi K$, despite the unclear situation in the SM:

The fit becomes more complicated than in the previous cases, because NP contributions with $\Delta I=1$ induce two new isospin amplitudes, 
corresponding to final $|K\pi\rangle$ states with $I=1/2$ or $I=3/2$.
Note that in this case, the contributions with $\Delta I=1$ are not expected to be suppressed.
In order to reduce the number of free parameters in the fit, and to avoid unphysical solutions, 
the following additional assumptions/approximations are applied:
Following the experimental observation, the direct CP asymmetry in the decay $B^-\to\pi^-\bar K^0$ is forced
to vanish identically, which yields the relation $r_0 \, e^{i\phi_{0}}= -r_1^{(1/2)} \, \, e^{i \phi_{1}^{(1/2)}} - r_1^{(3/2)} \,  e^{i \phi_{1}^{(3/2)}} \,.$
This effectively implies dealing with 
a $b \to s \bar{u}u$ operator which does not contribute to $B^- \to \pi^- \bar K^0$ in the naive factorization 
approximation. The amplitude parameters $\epsilon_{T,3/2} $ and $\phi_{T,3/2}$ are chosen to be equal and lie within the 
QCDF ranges, see \cite{FJM}. 
For $\theta_W=\pi-\gamma$, the fit results in
$r_1^{(1/2)}\in[0.02,\,0.08]$, $\phi_1^{(1/2)}\in[-2.84,\,-0.52]$, $r_1^{(3/2)}\in[0.00,\,0.23]$, and $\phi_1^{(3/2)}$ unconstrained, see also Table~\ref{tab:bestfit}. 
However, the QCDF input breaks reparametrization invariance, therefore the fit depends on $\theta_W$ in an essential way, see \cite{FJM,JPhd}.
Notably, also in this scenario the measured values for the time-dependent
CP asymmetry in $B\to\pi^0K_S$ are difficult to accomodate, as can be seen
in table~\ref{tab:bestfit}. With a phase differing strongly from the
SM one that is possible (for example with $\theta_W\sim\pi/3$), however only
with rather large NP contributions. While this is a first hint on a genuine 
NP phase, it is paid by the model-dependence mentioned before.
%
%
%
\section{Conclusions\label{Conclusions}}
The work presented here pursues a model-independent approach. Assuming the dominance of an individual NP operator, the analysis of 
$B \to J/\psi K$, $B \to \phi K$ and $B \to K \pi$ observables allows for infering semi-\-quan\-ti\-ta\-tive information about the relative 
size of NP contributions to $b \to s \bar{q}q$ operators. 
The main conclusions to be drawn are:
(i) All three modes discussed above prefer the inclusion of an operator transforming non-trivial under isospin, namely an 
operator with the structure $\mathcal{O}_{\rm NP}\sim(\bar{s}b)(\bar{u}u)$ provides a solution for all observed tensions.
(ii) From the comparison of isospin-averaged $B \to J/\psi K$ and $B \to \phi K$ observables it is found that --- after correcting
for relative penguin, phase-space and normalization factors --- NP contributions to $b \to s (c\bar c/\bar{s}s)$ operators
may be of similar size (order 10\% relative to a SM tree operator).
(iii) In all cases, in order to explain the tensions with SM expectations for CP asymmetries without fine-tuning of hadronic parameters, 
one has to require non-trivial weak phases ($\theta_W\neq 0,\pi$), 
which could be due to NP, albeit the case $\theta_W = \pi -\gamma_{\rm SM}$ is always allowed, too. A different weak phase is only
preferred in $B\to\pi K$, which is however only a very weak indication of a genuine NP phase. Consequently, these findings are still 
compatible with a SM scenario where non-factorizable QCD dynamics in matrix elements of subleading operators is unexpectedly large.

In the future, an improvement of experimental accuracy, in particular on the isospin-violating observables like the rate asymmetry, could lead to even more interesting constraints on the relative importance of different $b \to s q\bar q$ operators and their interpretation within
particular NP models.

\section*{Acknowledgements}
This work has been done in collaboration with Th. Feldmann and Th. Mannel.
It was supported by the EU MRTN-CT-2006-035482 (FLAVIAnet), by MICINN (Spain) under grant FPA2007-60323, and by the Spanish 
Consolider-Ingenio 2010 Programme CPAN (CSD2007-00042). 

\bibliography{ThesisBib}

\begin{thebibliography}{10}

\bibitem{FJM}
T.~Feldmann, M.~Jung, T.~Mannel.
\newblock \emph{JHEP} \textbf{08}, 066 (2008).

\bibitem{CKMfitter}
J.~Charles, et~al.
\newblock \emph{Eur. Phys. J.} \textbf{C41}, 1 (2005).
\newblock Updated results and plots available at: {\tt
  http://ckmfitter.in2p3.fr}.

\bibitem{FleischerMannel1}
R.~Fleischer, T.~Mannel.
\newblock \emph{Phys. Lett.} \textbf{B506}, 311 (2001).

\bibitem{Boos}
H.~Boos, T.~Mannel, J.~Reuter.
\newblock \emph{Phys. Rev.} \textbf{D70}, 036006 (2004).

\bibitem{Li}
H.-n. Li, S.~Mishima.
\newblock \emph{JHEP} \textbf{03}, 009 (2007).

\bibitem{GronauRosnersuppressedterms}
M.~Gronau, J.~L. Rosner.
\newblock \emph{Phys. Lett.} \textbf{B672}, 349 (2009).

\bibitem{Ciuchini2005}
M.~Ciuchini, M.~Pierini, L.~Silvestrini.
\newblock \emph{Phys. Rev. Lett.} \textbf{95}, 221804 (2005).

\bibitem{FFJM}
S.~Faller, M.~Jung, R.~Fleischer, T.~Mannel.
\newblock \emph{Phys. Rev.} \textbf{D79}, 014030 (2009).

\bibitem{HFAG}
E.~Barberio, et~al.
\newblock \emph{arXiv:} 0808.1297 (hep--ex) (2008).
\newblock Online update available at {\tt
  http://www.slac.stanford.edu/xorg/hfag}.

\bibitem{Gronauetal1}
M.~Gronau, et~al.
\newblock \emph{Phys. Rev.} \textbf{D52}, 6356 (1995).

\bibitem{JPhd}
M.~Jung.
\newblock Ph.D. thesis, Universit\"at Siegen (2009).
\newblock Http://dokumentix.ub.uni-siegen.de/opus/volltexte/2009/392/.

\bibitem{Grossmanetal}
Y.~Grossman, Z.~Ligeti, Y.~Nir, H.~Quinn.
\newblock \emph{Phys. Rev.} \textbf{D68}, 015004 (2003).

\bibitem{WilliamsonZupan}
A.~R. Williamson, J.~Zupan.
\newblock \emph{Phys. Rev.} \textbf{D74}, 014003 (2006).

\bibitem{Chengetal}
H.-Y. Cheng, C.-K. Chua, A.~Soni.
\newblock \emph{Phys. Rev.} \textbf{D72}, 014006 (2005).

\bibitem{Beneke}
M.~Beneke.
\newblock \emph{Phys. Lett.} \textbf{B620}, 143 (2005).

\bibitem{Neubert1998}
M.~Neubert.
\newblock \emph{JHEP} \textbf{02}, 014 (1999).

\bibitem{BBNS1}
M.~Beneke, et~al.
\newblock \emph{Nucl. Phys.} \textbf{B591}, 313 (2000).

\bibitem{Ciuchinietal2008}
M.~Ciuchini, et~al.
\newblock \emph{Phys. Lett.} \textbf{B674}, 197 (2009).

\bibitem{Fleischeretal2008}
R.~Fleischer, et~al.
\newblock \emph{Phys. Rev.} \textbf{D78}, 111501 (2008).

\bibitem{GronauRosner2008}
M.~Gronau, J.~L. Rosner.
\newblock \emph{Phys. Lett.} \textbf{B666}, 467 (2008).

\bibitem{Gronau2005}
M.~Gronau.
\newblock \emph{Phys. Lett.} \textbf{B627}, 82 (2005).

\bibitem{Baeketal2009}
S.~Baek, C.-W. Chiang, D.~London.
\newblock \emph{arXiv} 0903.3086 (hep--ph) (2009).

\end{thebibliography}
\end{document}